\documentstyle[12pt,epsfig]{article} 
\def\baselinestretch{1.3}
\newcommand{\ba}{\begin{array}}
\newcommand{\ea}{\end{array}}
\newcommand{\bd}{\begin{displaymath}}
\newcommand{\ed}{\end{displaymath}}
\newcommand{\be}{\begin{equation}}
\newcommand{\ee}{\end{equation}}
\newcommand{\bea}{\begin{eqnarray}}
\newcommand{\eea}{\end{eqnarray}}

\def\ve{\varepsilon}

\def\q2 {q^2}

\def\bt{\begin{table}}
\def\et{\end{table}}
\catcode`@=11 
\def \gsim{\mathrel{\mathpalette\@versim>}}
\def \lsim{\mathrel{\mathpalette\@versim<}}
\def \@versim#1#2{\lower0.4ex\vbox{\baselineskip\z@skip\lineskip\z@skip
     \lineskiplimit\z@\ialign{$\m@th#1\hfil##\hfil$%
     \crcr#2\crcr\sim\crcr}}}
\catcode`@=12 
\begin{document}
%
%
\begin{center}

{\large\bf Unitarity violation in sequential neutrino mixing in a
model of extra dimensions}\\[15mm]

Subhaditya Bhattacharya\footnote{E-mail: subha@mri.ernet.in}, Paramita Dey
\footnote{E-mail: paramita@mri.ernet.in}
and Biswarup Mukhopadhyaya\footnote{E-mail: biswarup@mri.ernet.in}\\
{\em Regional Centre for Accelerator-based Particle Physics \\
     Harish-Chandra Research Institute\\
Chhatnag Road, Jhunsi, Allahabad - 211 019, India}
\\[20mm] 
\end{center}

\begin{abstract} 
We investigate the possibility of unitarity violation in the
sequential neutrino mixing matrix in a scenario with extra compact
spacelike dimensions. Gauge singlet neutrinos are assumed to propagate
in one extra dimension, giving rise to an infinite tower of states in
the effective four-dimensional theory. It is shown that this leads to
small lepton-number violating entries in the neutrino mass matrix,
which can violate unitarity on the order of one per cent.
\end{abstract}

\vskip 1 true cm

\newpage
\setcounter{footnote}{0}

\def\baselinestretch{1.5}
\section{Introduction}

The ever-consolidating evidence in favour of neutrino masses and
mixing has spawned a large volume of speculations on new physics
possibilities that could be at their origin. Considering the three
light sequential neutrinos, many proposed scenarios, including seesaw
models of type I \cite{type1}, II \cite{type2} or III \cite{type3},
ensure unitarity to a high degree of precision in the
Pontecorvo-Maki-Nakagawa-Sakata (PMNS) matrix describing mixing in the
lepton sector \cite{pmns}. A measured departure from such unitarity,
evinced from precision data in the neutrino sector, may thus point
towards some novel mechanism for the generation of neutrino
masses. One such possibility arises when at least one small gauge
singlet Majorana mass term enters into an extended neutrino mass
matrix. Under certain conditions, this situation passes off as
`inverse seesaw mechanism' \cite{ma1,ma2,others}. It has been
demonstrated in a number of recent works that this can lead to a
violation of unitarity at the level of about 1\% or more in the
$3\times 3$ light sequential neutrino mass matrix, due to mixing with
additional sterile states \cite{ma1,ma2,others,werner}. A pertinent
question to ask is: does such a situation fit into some of the popular
scenarios of new physics at the TeV scale?

The experimental constraints on the loss of unitarity as well as its
testability in neutrino oscillation experiments has been investigated
recently \cite{werner,antus}. As for theoretical models, a
GUT-inspired scenario, based on $SO(10)$ with a breaking chain
involving an extra $U(1)$ gauge symmetry surviving at low scale, has
been considered recently for this purpose \cite{ma2}. This scenario
has been shown to lead to two-loop generation of some small Majorana
masses and consequently lead to the inverse seesaw mechanism. It has
also been suggested that a supersymmetric model including two types of
gauge singlet neutrino superfields may produce effects of this kind
\cite{susy}. In a number of other model-building ventures, too, the
effect mentioned above emerges as a consequence \cite{etc}.

Phenomenological implications of unitarity violation in the PMNS
matrix, including its signatures in phenomena driven by neutrino
oscillation, have been recently investigated \cite{antus}. In this
paper, we point out that a loss of unitarity in the PMNS matrix can
also arise if one has extra flat spacelike dimensions, with gauge
singlet neutrinos propagating in {\it one} extra dimension, and
lending small diagonal elements to an extended neutrino mass matrix.

Mechanisms of neutrino mass generation have been frequently suggested
in models of compact extra spacelike dimensions, both flat
\cite{DDG2,ADDM,AP1,IP,MM,allothers,Bhattacharyya:2002vf} and warped
\cite{PDs}. Here we consider a minimal higher-dimensional framework
where the standard model (SM) fields all lie on a 3-brane, while one
or more gauge singlet neutrino propagate along {\em one flat extra
dimension} \cite{DDG2,ADDM}. However, there can in principle be
several extra spacelike dimensions where gravity propagates, thereby
evading the already established lower limits on the number of such
dimensions \cite{gravexp}. An orbifold symmetry is further imposed
along the compact direction containing the neutrino(s), so that one
obtains only one (right-handed) chirality for the $n=0$ Kaluza-Klein
(KK) mode. It has been demonstrated earlier that this scenario can
naturally suppress neutrino masses via a Type I \cite{type1} seesaw
mechanism.

The gauge singlet neutrinos can have Majorana masses in five
dimensions to start with.  We are specially interested in the
situation where a strong cancellation between this mass and the KK
tower mass leads to a very small entry in the effective neutrino mass
matrix in four dimensions. We show that the resulting mass matrix has
additional `sterile' states mixing appreciably with the sequential
neutrinos. It is found that one can consequently expect the violation
of unitarity in the $3\times 3$ (PMNS) matrix in certain regions of
the parameter space of such a model.
 
In Section 2, we outline some scenarios that lead to loss of unitarity
of the PMNS matrix, including the inverse seesaw mechanism. The
extra-dimensional model under investigation is briefly reviewed in
Section 3. The viability of a substantial loss of PMNS unitarity is
numerically demonstrated in Section 4.  We summarise and conclude in
Section 5.

\section{Loss of unitarity in the PMNS matrix}

In general, the well-known Type I seesaw mechanism involving a light
and a heavy neutrino also involves a departure from unitarity in the
PMNS matrix. However, this departure is immeasurably tiny, since the
seesaw mass scale in invariably much higher than the light neutrino
masses \cite{type1}. An exception to this may occur if a very small
Majorana mass is introduced. However, this choice is inhibited by (a)
the need of justifying such a small $\Delta L = 2$ mass in terms of
new physics, and (b) the need of introducing excessively suppressed
Dirac masses for generating light neutrinos, which essentially
destroys the motivation of the seesaw mechanism.

The situation can be different when one has more than one
two-component sterile neutrinos. It has been shown in a number of
recent works \cite{ma1,ma2,others,werner} that this allows one to
insert small lepton-number violating mass terms in diagonal positions
of the neutrino mass matrix, and the off-diagonal entries need not all
be much smaller. Its most noticeable consequence is a loss of
unitarity at the level of 1\% or more in the $3\times 3$ (PMNS) part
of the neutrino mass matrix.

Several kinds of scenarios that meet this description are found in the
literature \cite{ma1,others,werner}. Here we mention two classes only
among these, considering for illustration one sequential and two
sterile species in each case.  The first \cite{ma1} is one of the form
\be
\label{ema}
{\cal M}\ =\ \left(\! \begin{array}{ccc}
0 & m_{D} & 0 \\
m_D & m_R & m_N  \\
0 & m_N & m_L 
\end{array}\!\right)\,,
\ee
\noindent
in the basis (${\bar{\nu}}_L, N_R, {\bar{N}}_L$), where the last two
are gauge singlets. The masses $m_{L,R}$ arise from $\Delta L =2$
terms.  For $m_{L,R} << m_D, m_N$, this not only yields an active
neutrino mass eigenstate in the right order, but also leads to
active-sterile mixing at the level of 1\% for appropriate choice of
the mass parameters (say, for example, $m_D \sim 10$ GeV, $m_N \sim 1$
TeV, and $m_L \sim 10$ keV). A corresponding situation with three
sequential neutrinos will show unitarity violation at the same level
in the PMNS matrix, but with an additional light sterile neutrino.
Since the light (sequential) neutrino mass vanishes in the limit $m_L
\rightarrow 0$, it is often called an `inverse seesaw' scenario.

Another situation that one can consider has the same choice of
neutrino basis states, but a mass matrix of the form \cite{werner}
\be
\label{werner}
{\cal M}\ =\ \left(\! \begin{array}{ccc}
0 & m_{D} & m_N \\
m_D & m_R & 0  \\
m_N & 0 & m_L 
\end{array}\!\right)\,,
\ee
\noindent
with $m_D<<m_R$ and $m^2_N/m_R<<m_L<<m_N<<m_R$. It has been found that
this situation, too, leads to light sequential neutrino(s) and
unitarity violation at the same level ($\sim 1$ \%), for appropriate
choice of parameters (say, for example, $m_D, m_N \sim 1$ MeV, $m_R
\sim 1$ TeV and $m_L \sim 100$ eV). The difference with the previous
situation is that (a) one obtains a light sterile neutrino even with
one sequential family, and (b) the sequential neutrino mass does not
vanish in the limit $m_L \rightarrow 0$. This makes it deviate from an
inverse seesaw scenario in the strict sense, although it is equally
interesting from the viewpoint of unitarity loss of the PMNS
matrix. Since such unitarity loss is a very interesting consequence
that is experimentally testable, it is worth exploring if it occurs in
some otherwise well-motivated theories beyond the SM.  In the next two
sections we outline one such scenario, and go on to examine its
potential for generating unitarity loss.

\section{A model with extra dimensions}

In this section we describe the model adopted for illustrating our
point. It assumes extra flat compact spacelike dimensions where
gravity can propagate. The SM fields are confined to a 3-brane which
constitutes a `slice' in the higher-dimensional space. So far it is
very similar to the Arkani Hamed-Dimopoulos-Dvali (ADD) scenario
\cite{flatgeo}, excepting that it includes an effort to account for
neutrino masses, through the introduction of one gauge singlet
neutrino propagating in {\it one extra dimension only}
\cite{DDG2,ADDM,AP1,IP,MM,allothers}. Thus, while all the
phenomenology involving gravitons remains similar to that in the ADD
framework with several extra dimensions, one can consider just the
five-dimensional subspace for studying neutrino physics. We make our
analysis simple by adhering to one generation of SM neutrinos. The
fifth flat dimension ($y$), along which propagates the right handed
neutrino ($N(x,y)$), is compactified over an $S^1/Z_2$ orbifold where
$R$ is the radius of compactification. The preservation of the $Z_2$
invariance necessitates the existence of at least two symmetrically
placed branes, and the SM fields lie on either of them. Thus the
complete leptonic field content of the model is
\begin{equation}
\label{LSM}
L(x)\ =\ \left( \begin{array}{c} \nu_\ell (x) \\ \ell_L (x)
\end{array} \right) ,\qquad \ell_R (x)\,, \qquad N(x,y)\ =\ \left(
\begin{array}{c} \xi (x,y) \\ \bar{\eta} (x,y) \end{array} \right)\,,
\end{equation}
where $\nu_\ell$, $\ell_L$, $\ell_R$ are Weyl spinors in four
dimensions, and $\xi$, $\eta$ are two-component spinors in five
dimensions.  Under $S^1/Z_2$, the latter may be associated with
opposite parities:
\begin{equation}
\label{yparity}
\xi (x,y)\ =\ \xi (x,-y)\,,\qquad  \eta (x,y)\ =\ - \eta (x,-y)\,.
\end{equation}
The brane where the SM is localised, can be assumed to be at $y=a$
just for generality, instead of at the orbifold fixed point $y=0$.  We
shall see later that this adds to the freedom of the model. The
generic effective four-dimensional Lagrangian of this model is given
by,
\begin{eqnarray}
\label{Leff} 
{\cal L}_{\rm eff} & =& \int\limits_0^{2\pi R}\!\! dy\ \bigg\{\,
\bar{N} \Big( i\gamma^\mu \partial_\mu\, +\, \gamma_5 \partial_y \Big)
N\ -\ \frac{1}{2}\,\Big( M N^T C^{(5)-1} \gamma_5 N\ +\ {\rm h.c.}
\Big) \nonumber\\ &&+\,\delta (y-a)\, \bigg[\,
\frac{{h}_1}{(M_F)^{1/2}}\, L\tilde{\Phi}^* \xi\, +\,
\frac{{h}_2}{(M_F)^{1/2}}\, L \tilde{\Phi}^* \eta\ +\ {\rm
h.c.}\,\bigg]\ +\ \delta (y-a)\, {\cal L}_{\rm SM}\, \bigg\}\, ,\quad
\end{eqnarray}
where $\tilde{\Phi} = i\sigma_2 \Phi^*$, ${\cal L}_{\rm SM}$ is the SM
Lagrangian, $M$ is the Majorana mass for $N$ (we do not specify its
scale for the moment), $C^{(5)}$ is the 5-dimensional charge
conjugation operator and $M_F$ is the fundamental gravity scale.  The
Yukawa couplings in five dimensions, $h_{1,2}$, are assumed to be
${\cal{O}}(1)$. For gravity propagating in a $d$-dimensional bulk,
\bea
\label{cutoff}
M_P=(2\pi M_F R)^{d/2}M_F,  
\eea 
for the simple case where all the compactification radii are of equal
size $R$, $M_P$ being the four-dimensional Planck scale. A Dirac mass
term $m_D \bar{N}N$ is not allowed in equation (\ref{Leff}) because of
the $Z_2$ symmetry. \newline Following equation (\ref{yparity}), the
two-component spinors $\xi$ and $\eta$ can be expanded as,
\begin{eqnarray}
\label{xi}
\xi (x,y) &=& \frac{1}{\sqrt{2\pi R}}\ \xi_0 (x)\ +\ 
\frac{1}{\sqrt{\pi R}}\ \sum_{n=1}^\infty\, \xi_n (x)\ 
\cos\bigg(\,\frac{ny}{R}\,\bigg)\,,\\
\label{eta}
\eta (x,y) & =& \frac{1}{\sqrt{\pi R}}\ \sum_{n=1}^\infty\, \eta_n (x)\ 
\sin\bigg(\,\frac{ny}{R}\,\bigg)\,,
\end{eqnarray}
where the chiral spinors $\xi_n(x)$ and $\eta_n(x)$ form an infinite
tower of KK fields. Using these expansions and integrating out the
$y$-coordinate, the effective Lagrangian reduces to
\begin{eqnarray}
\label{Leff1KK}
{\cal L}_{\rm eff} & = & {\cal L}_{\rm SM}\ +\ \bar{\xi}_0
( i\bar{\sigma}^\mu \partial_\mu) \xi_0\ 
+\ \Big(\, \bar{h}^{(0)}_1\, L\tilde{\Phi}^* \xi_0\ -\
\frac{1}{2}\, M\, \xi_0\xi_0\ +\ {\rm h.c.}\,\Big)\
 +\ \sum_{n=1}^\infty\, \bigg[\, \bar{\xi}_n 
( i\bar{\sigma}^\mu \partial_\mu) \xi_n\nonumber\\
&& +\, \bar{\eta}_n ( i\bar{\sigma}^\mu \partial_\mu) \eta_n\
+\ \frac{n}{R}\, \Big( \xi_n \eta_n\, +\, \bar{\xi}_n
\bar{\eta}_n\Big) -\ \frac{1}{2}\, M\, 
\Big( \xi_n\xi_n\, +\, \bar{\eta}_n\bar{\eta}_n\
+\ {\rm h.c.}\Big)\nonumber\\
&& +\, \sqrt{2}\, \Big(\, \bar{h}^{(n)}_1\, L\tilde{\Phi}^* \xi_n\ +\
\bar{h}^{(n)}_2\, L\tilde{\Phi}^* \eta_n\ +\ {\rm h.c.}\,\Big)\, \bigg]\, 
\end{eqnarray} 
in a basis in which $M$ is positive, and with:
\begin{eqnarray}
\label{h1n}
\bar{h}^{(n)}_1 &=& \frac{h_1}{(2\pi M_F R)^{1/2}}\ 
\cos \bigg(\,\frac{na}{R}\,\bigg)\ =\ \bigg(\,\frac{M_F}{M_{\rm
P}}\,\bigg)^{1/d}\  
h_1 \cos \bigg(\,\frac{na}{R}\,\bigg)\ =\ 
\bar h_1 \cos \bigg(\,\frac{na}{R}\,\bigg)\,,\\
\label{h2n}
\bar{h}^{(n)}_2 &=& \frac{h_2}{(2\pi M_F R)^{1/2}}\ 
\sin \bigg(\,\frac{na}{R}\,\bigg)\ =\ \bigg(\,\frac{M_F}{M_{\rm
P}}\,\bigg)^{1/d}\  
h_2 \sin \bigg(\,\frac{na}{R}\,\bigg)\ =\
\bar h_2 \sin \bigg(\,\frac{na}{R}\,\bigg)\, .
\end{eqnarray}
For deriving the last two equalities on the right hand sides of
equations (\ref{h1n})-(\ref{h2n}), we have made use of equation
(\ref{cutoff}).

Equations (\ref{h1n}) and equation (\ref{h2n}) imply that the induced
four-dimensional Yukawa couplings $\bar{h}^{(n)}_{1,2}$ can get
suppressed by many orders depending on the hierarchy between $M_P$ and
$M_F$; for example, if gravity and the bulk neutrino feel the same
number of extra dimensions, say $d=1$, then these couplings are
suppressed by a factor $M_F/M_{\rm P} \sim 10^{-15}$, for $M_F \approx
10$~TeV (see also \cite{DDG2,ADDM}).

It is clear from equation (\ref{LSM}) that $\xi$ and $\bar{\eta}$ have
the same lepton number. Thus, the simultaneous presence of the two
operators $L\tilde{\Phi}^*\xi$ and $L\tilde{\Phi}^*\eta$ in equation
(\ref{Leff1KK}) leads to lepton number violation. Such coexistence of
the two operators is possible only if we allow the brane to be shifted
by an amount $a (\ne 0)$ from the orbifold fixed points ($y=0$, $\pi
R$). Such a shifting of the brane, respecting the $Z_2$ invariance of
the original higher dimensional Lagrangian, has been shown to be
possible under certain restrictions in Type-I string theories
\cite{GP}. As indicated in \cite{DDG2,Bhattacharyya:2002vf}, the $Z_2$
invariance can be taken care of by allowing the replacements
\begin{eqnarray}
\label{exp1}
\xi\, \delta (y - a) &\to & \frac{1}{2}\; \xi\,\Big[\, \delta ( y -
a)\: +\: \delta ( y + a - 2\pi R) \, \Big]\,,\nonumber\\ \eta\, \delta
(y - a) &\to & \frac{1}{2}\; \eta\,\Big[\, \delta ( y - a)\: -\:
\delta ( y + a - 2\pi R) \, \Big]\,,
\end{eqnarray}
with $0\le a < \pi R$ and $0\le y \le 2\pi R$. Here we re-iterate that
a $Z_2$-invariant implementation of brane-shifted couplings requires
the existence of at least two branes placed at $y= a$ and $y = 2\pi R
-a$.

A remarkable feature of the brane-shifted framework was pointed out in
\cite{Bhattacharyya:2002vf}, where it has been shown that in such a
framework it is possible to completely decouple the effective
Majorana-neutrino mass $\langle m\rangle$ and the scale of light
neutrino masses, so as to have $\langle m\rangle$ within an observable
range. Therefore, the Lagrangian (\ref{Leff1KK}) contains two types of
Majorana neutrino mass terms (involving respectively the parameters
$M$ and $\bar h_2^{(n)}$) both of which lead to a breaking of
$L$. Such $L$-breaking is a necessary ingredient of leptogenesis.

Following the notations of reference \cite{DDG2}, we now introduce the
weak basis for the KK Weyl-spinors, by defining
\begin{equation}
\label{xieta}
\chi_{\pm n}\ =\ \frac{1}{\sqrt{2}}\, (\,\xi_n\: \pm\:
\eta_n\,),
\end{equation}
followed by a rearrangement of the states $\xi_0$ and $\chi^\pm_n$, such
that, for a given value of $n$ (say, $n=k_0$), the smallest diagonal
entry of the neutrino mass matrix is
\bea
\label{vareps}
\varepsilon = {\rm min}\, \Big( \left| M - \frac{k_0}{R} \right| \Big)
\leq 1/(2R).  \eea
After re-ordering, we can define the multiplet $\Psi_\nu$ consisting
of the Majorana spinors
\begin{equation}
\label{Psinu}
\Psi^T_\nu \ =\ \left[\, 
\left(\! \begin{array}{c} \chi_{\nu_{\ell}} \\ \bar{\chi}_{\nu_{\ell}}
\end{array}\!\right)\,,\ 
\left(\! \begin{array}{c} \chi_{k_0} \\ \bar{\chi}_{k_0} 
\end{array}\!\right)\,,\
\left(\! \begin{array}{c} \chi_{k_0-1} \\ \bar{\chi}_{k_0-1} 
\end{array}\!\right)\,,\
\left(\! \begin{array}{c} \chi_{k_0+1} \\ \bar{\chi}_{k_0+1}
\end{array}\!\right)\,,\
\cdots\,,
\left(\! \begin{array}{c} \chi_{k_0-n} \\ \bar{\chi}_{k_0-n} 
\end{array}\!\right)\,,\
\left(\! \begin{array}{c} \chi_{k_0+n} \\ \bar{\chi}_{k_0+n} 
\end{array}\!\right)\,,\
\cdots\ \right]
\end{equation}
while the effective Lagrangian for right handed neutrinos reduces to
\begin{equation}
\label{Lkinorb}
{\cal L}_{\rm kin}\ =\ \frac{1}{2}\, \bar{\Psi}_\nu\,\Big(\, 
i\!\not\!\partial\ -\ {\cal M}^{\rm KK}_\nu\, \Big)\, \Psi_\nu\,,
\end{equation}
where ${\cal M}^{\rm KK}_\nu$ is the corresponding neutrino mass
matrix given by
\begin{equation}
\label{Morbshift}
{\cal M}^{\rm KK}_\nu\ =\ \left(\! \begin{array}{ccccccc}
0 & m^{(0)} & m^{(-1)} & m^{(1)} & m^{(-2)} & m^{(2)} & \cdots \\
m^{(0)} & \varepsilon & 0 & 0 & 0 & 0 & \cdots  \\
m^{(-1)} & 0 & \varepsilon - \frac{1}{R} & 0 & 0 & 0 & \cdots \\
m^{(1)} & 0 & 0 & \varepsilon + \frac{1}{R} & 0 & 0 & \cdots \\
m^{(-2)} & 0 & 0 & 0 & \varepsilon - \frac{2}{R} & 0 & \cdots \\
m^{(2)} & 0 & 0 & 0 & 0 & \varepsilon + \frac{2}{R} & \cdots \\
\vdots & \vdots & \vdots & \vdots & \vdots & \vdots & \ddots
\end{array}\!\right)\,.
\end{equation}
The most important consequence of such a rearrangement is that the
mass scale $M$, which we did not specify earlier but which could be
arbitrarily large, is now replaced by the light mass scale
$\varepsilon$.  The entries in the first row and the first column of
${\cal M}^{\rm KK}_\nu$ are given by the relation,
\begin{eqnarray}
\label{mk0}
m^{(n)} &=& \frac{v}{\sqrt{2}}\, \bigg[\, \bar{h}_1\, 
\cos\bigg(\frac{ (n-k_0) a}{R}\bigg)\: +\: \bar{h}_2\, 
\sin\bigg(\frac{ (n-k_0) a}{R}\bigg)\,\bigg]\ =\
m\,\cos\bigg(\frac{na}{R}\, -\, \phi_h\,\bigg)\,,\qquad
\end{eqnarray}
with
\bea 
\label{initialy}
m &=& \frac{v}{2}\sqrt{\frac{h^2_1 + h^2_2}{\pi M_F R}} =
\frac{m_{\rm max}}{\sqrt{M_FR}}, \\
\label{phih}
\phi_h &=& \tan^{-1} \left(\frac{h_2}{h_1}\right) + k_0 \frac{a}{R},  
\eea
where $v$ is the vacuum expectation value of the SM Higgs boson.

\section{Unitarity loss with extra dimensions: some numerical
  illustrations} 

Here we show that a substantial loss of unitarity of the PMNS matrix
can occur in different allowed regions of the parameter space of the
model described in the previous section. The first issue is, of
course, insuring at least one small entry in diagonal positions of the
neutrino mass matrix ${{\cal M}^{\rm KK}_\nu}$.  Equation
(\ref{vareps}) tells us that $\ve \le 1/2R$. There is no other
theoretical or phenomenological constraint on $\ve$. Thus $\ve$
qualifies to be the small diagonal element which can be potentially
responsible for a departure from unitarity.

$M_F$, the Planck mass in five dimensions, is expected to be $\gsim$
TeV, since gravitational effects will otherwise become important in
low-energy physics. At the same time, in order to ensure that physics
along the compact dimension(s) is not plagued with trans-Planckian
effects, one should have $1/R \le M_F$. Thus, in the expression for
$m$ in equation (\ref{initialy}), $M_F R$ is at least of order
unity. Given the fact that the five-dimensional Yukawa couplings
$h_{1,2}$, too, are {\it prima facie} of the order of unity, this
implies that $m$ can at most be around $v$, the electroweak symmetry
breaking scale.

We indicated two scenarios of unitarity loss in section 2: ({\it i})
where the off-diagonal elements in the extended neutrino mass matrix
are all larger than the diagonal ones, and ({\it ii}) where diagonal
elements excepting the smallest one are larger than the off-diagonal
ones. Let us first examine whether the extra-dimensional model under
scrutiny answers to both of these scenarios.

The first possibility demands 
\be
m > {\frac{1}{R}}
\label{p1}
\ee
since, with $a\neq 0$, $m^{(n)}$ can approach $m$ for some value of
$n$ along the tower. Using equation (\ref{initialy}) in (\ref{p1}),
one obtains
\be
\frac{m^2_{\rm max}}{M_F} > {\frac{1}{R}}
\label{p2}
\ee
The inequality should hold for the maximum value of the right-hand
side for a given $M_F$, which is $M_F$ itself. Thus we have
\be
{M_F} < m_{\rm max} 
\label{p3}
\ee
Therefore, demanding $m > 1/R$ implies that the five-dimensional
Planck scale has to be brought down below $ m_{\rm max}$ which is just
about the electroweak symmetry breaking scale, and hence is
inadmissible.
  
The first scenario is thus disfavoured in this model. On the other
hand, since the diagonal elements ${{\cal M}^{\rm KK}_\nu}(i,i),~i\geq
3$ are always greater in magnitude than $m^{(n)}$ for all values of
$n$, one can say that for sufficiently small $\ve$, this model
provides an opportunity for unitarity loss in the PMNS matrix in the
sense of the second scenario mentioned in Section 2.

The value of $1/R$, on the other hand, is not subject to any general
constraint stronger than that arising from the validity of Newton's
law of gravitation down to about $10^{-2} mm$, which essentially
allows $1/R$ to have any value $\gsim 10^{-8}$ MeV \cite{gravexp}.
Precision electroweak constraints do not tighten the constraint, since
the tower resulting from the compactification of the extra dimension
corresponds to gauge singlet neutrinos only. Thus, in order to have
loss of unitarity in the PMNS matrix, we are faced with two
possibilities, namely, (a) $\ve << 1/R$, and (b) $\ve \simeq 1/R$. We
show below that both of these situations are possible.

Our principal aim is to check if it is possible to have the sequential
neutrino masses in the right order ($\sim 10^{-2}$ eV), and at the
same time have substantial violation of unitarity.  The latter
requires that the squares of elements of some particular column in the
full mixing matrix beyond the PMNS block add up to ${\cal O}
(10^{-4})$. This sum is defined as $\delta^2$ here. \footnote{In
references \cite{werner,antus}, the violation of unitarity has been
defined in terms of a parameter $\eta$. It is easy to check that
$2\eta~=~\delta^2$.}

We carry out this investigation in the simplified situation, with one
sequential neutrino flavour and just one gauge singlet neutrino in the
bulk. We shall comment later on the generalisations necessary to
generate the actual pattern of masses and mixing. As far as the
violation of PMNS unitarity is concerned, however, the conclusions we
reach below remain valid even when such generalisations are made.

{\underline{\bf Case (a)}}: This implies a fine cancellation between
the bulk mass $M$ and some integral multiple of $1/R$. While a
dynamical explanation of this is difficult, it is not entirely
unlikely, as both $M$ and $1/R$ can rather naturally be around the
TeV-scale, and there is a distinct possibility of the two of them
having near-coincident values.

A few illustrative points in the parameter space for this case are
shown in Table \ref{tab:1}. We have confined ourselves to 1 MeV $\leq
1/R \leq$ 10 TeV. It is found that, to get a substantial unitarity
violation ($\delta \geq$ 0.5 \%) and neutrino mass in the right order,
the largest possible value of $\ve$ that we can take is $\simeq
10^{-6}$. In this case, the $2\times2$ block in the upper left corner
of the neutrino mass matrix effectively determines the masses of the
sequential and the lightest sterile neutrino, given as
$({m^{(0)}})^2/\ve$ and $\ve$ respectively. One further has:
\be 
\label{expl}
\delta \sim \left[\frac{({m^{(0)}})^2}{\ve}\right] \frac{1}{m^{(0)}}
\ee
Therefore if $\ve$ is increased, the concomitant enhancement in
$m^{(0)}$, required to keep the sequential neutrino mass unaffected,
ends up suppressing $\delta$. The values of $m^{(0)}$ required point
towards $\phi_{h} \simeq \pi/2$. On the other hand, the fact that
$m^{(1)}$ can vary over a wide range, implies that the brane-shift
parameter $a$ can vary from zero to $0.1 R$ approximately. It should
be noted that Table 1 includes one sample corresponding to
$m^{(0)}=m^{(1)}$, which means $a=$0. Thus, in this case, large
unitarity violation is consistent with both the cases where the brane
is located at orbifold fixed point and where it is noticeably shifted.
\begin{table}[htb]
\begin{center}
\begin{tabular}{||c|c|c|c|c||}
\hline
\hline
{\bf $1/R$}& {\bf $\ve$}&{\bf $m^{(0)}$}&{\bf $m^{(-1)}=m^{(+1)}$}& 
{\bf $\delta$ (\%)} \\
\hline
\hline
 $10,000$ & $10^{-7}$ & $10^{-9}$ & 100 & 1.7 \\
\hline
      &  $10^{-9}$ & $10^{-10}$ & 0.001 & 0.99 \\
      &  $10^{-9}$ & $10^{-7}$ & 0.001 & 1.0 \\
      &  $10^{-8}$ & $3\times 10^{-10}$ & 0.001 & 3.3 \\
      &  $10^{-7}$ & $10^{-9}$ & 0.001 & 1.0 \\
      &  $10^{-7}$ & $10^{-9}$ & 0.005 & 1.0 \\
1000  &  $10^{-7}$ & $10^{-9}$ & 0.01 & 1.0 \\
      &  $10^{-7}$ & $10^{-9}$ & 0.1 & 1.0 \\
      &  $10^{-7}$ & $10^{-9}$ & 1.0 & 1.0 \\
      &  $10^{-7}$ & $10^{-9}$ & 10.0 & 1.7 \\
      &  $10^{-7}$ & $10^{-9}$ & 20.0 & 3.0 \\
      &  $10^{-7}$ & $10^{-9}$ & 25.0 & 3.6 \\
\hline
1000  &  $3\times10^{-8}$ & $3\times10^{-10}$ & $3\times10^{-10}$
& 1.0 \\
\hline
10    &  $10^{-7}$ & $10^{-9}$ & 0.001 & 1.0 \\
\hline
      &  $10^{-7}$ & $10^{-9}$ & 0.001 & 1.0 \\
1     &  $10^{-6}$ & $3\times10^{-9}$ & 0.005 & 0.8 \\
      &  $10^{-6}$ & 6 $\times10^{-9}$ & 0.005 & 0.9 \\
\hline
0.01  &  $10^{-7}$ & $10^{-9}$ & $10^{5}$ & 1.2 \\
\hline
0.001  &  $10^{-7}$ & $10^{-9}$ & $10^{5}$ & 1.7 \\
\hline
\hline
\end{tabular}\\
\caption {\small \it {Different sample points in the parameter space
    of the model where substantial unitarity violation takes place,
    for $\ve << 1/R$. All mass parameters are in GeV.}}
\label{tab:1}       
\end{center}
\end{table}

{\underline{\bf Case (b)}}: In this case $\ve$ and $1/R$ can be
relatively close to each other. Therefore, no drastic cancellation is
required between them, and no allegation of fine-tuning can be
levelled against such a scenario.

Some sample results for this case are presented in Table
\ref{tab:2}. It should be noted that values of $1/R$ of comparable
smallness as that of $\ve$ implies that the part of ${\cal M}^{\rm
KK}_\nu$ beyond the upper left $2\times 2$ block no longer tends to
decouple. An immediate consequence is that $m^{(\pm1)}$ have to be as
small as $m^{(0)}$ when $1/R$ is small. This in turn drives the value
of the brane-shift parameter $a$ to very small values. Therefore,
unlike in the previous case, the brane is compelled to be close to the
orbifold fixed points.

The figures included in Table \ref{tab:2} are self-explanatory; the
provision for substantial unitarity violation is clearly
there. However, there is a relative paucity of available points
compared to the previous case. This is because the part of ${\cal
M}^{\rm KK}_\nu$ beyond the $2\times 2$ block is not ineffectual in
determining the sequential neutrino mass and its mixing with light
sterile states. In order to comply with all constraints there, one
therefore requires a correlation between $\ve$ and $1/R$, in contrast
to the situation with $\ve<<1/R$, thereby restricting the allowed
points in the parameter space.

The numerical results presented by us are obtained through the
diagonalisation of the $4\times 4$ neutrino mass matrix, including a
tower of states up to the first KK excitation only.  We have, however,
checked that the results do not change qualitatively upon the
inclusion of additional towers and the resulting augmentation of the
mass matrix.

\begin{table}[htb]
\begin{center}
\begin{tabular}{||c|c|c|c|c||}
\hline
\hline
{\bf $1/R$}& {\bf $\ve$}&{\bf $m^{(0)}$}&{\bf $m^{(-1)}=m^{(+1)}$}& 
{\bf $\delta$ (\%)} \\
\hline
\hline
            & $10^{-6}$ & $10^{-8}$ & $3.5 \times 10^{-8}$ & 1.0 \\
$10^{-5}$ & $2 \times 10^{-6}$ & $10^{-8}$ & $3.5 \times
10^{-8}$ & 0.7 \\
            & $5 \times 10^{-6}$ & $10^{-8}$ & $3.0 \times
10^{-8}$ & 0.7 \\
\hline
$10^{-5}$ & $5 \times 10^{-6}$ & $3 \times 10^{-8}$ & $3 \times
10^{-8}$ & 0.9 \\

\hline
$10^{-6}$ & $5 \times 10^{-7}$ & $3 \times 10^{-9}$ & $10^{-8}$ 
& 2.2 \\
\hline
\hline
\end{tabular}\\
\caption {\small \it {Different sample points in the parameter space
    of the model where substantial unitarity violation takes place,
    for $\ve \simeq 1/R$. All mass parameters are in GeV.}}
\label{tab:2}       
\end{center}
\end{table}

\section{Summary and conclusions}

We have studied a popular model of flat extra compact spacelike
dimensions. A gauge singlet neutrino is assumed to propagate in one
extra dimension. The `bulk' mass possessed by this neutrino can
undergo cancellation with the KK tower mass for some member of the
tower, giving rise to at least one small diagonal entry ($\ve$) in the
infinite-dimensional neutrino mass matrix in four dimensions.  We show
that this can cause substantial mixing between the sequential and
sterile neutrinos without violating any existing constraint. The
consequence is a departure from unitarity of the PMNS matrix, both for
$\ve << 1/R$ and $\ve \simeq 1/R$, $R$ being the radius of the compact
dimension housing the gauge singlet neutrino.

For the sake of simplicity, we have presented our results for one
sequential neutrino. It can be easily checked that the conclusions are
valid with additional generations. In fact, the constraints on
$\delta$ are easy to satisfy, since the strongest constraint on the
PMNS matrix is on its $(1,2)$'{\it th} element
\cite{werner,antus}. The relatively unconstrained mixing of, for
example, $\nu_\tau$ with a sterile neutrino can accommodate the values
of $\delta$ obtained here. On the other hand, it may be difficult to
accommodate the neutrino mixing data and mass hierarchies with one
sterile bulk neutrino only \cite{neudata}. At least two such neutrinos
can, however, accommodate everything rather easily, thanks to the
additional Yukawa couplings available, which are essentially free
parameters. Our general conclusions are unaffected by such extensions.

In conclusion, the phenomenon of unitarity violation in the PMNS
matrix can be motivated rather well in a model of extra dimensions.
This brings to the fore the likely connection between subtleties of
the neutrino sector and theories which advocate strikingly new physics
around the Tev scale.

{\bf Acknowledgment:} This work was partially supported by funding
available from the Department of Atomic Energy, Government of India
for the Regional Centre for Accelerator-based Particle Physics,
Harish-Chandra Research Institute.


\end{document}